\documentclass[twocolumn,nofootinbib, superscriptaddress, preprintnumbers]{revtex4}
\usepackage{amsmath}
\usepackage{graphicx}
\usepackage{color}

\newcommand{\Smax}{\ensuremath{S_{ij}^\mathrm{max}}}
\newcommand{\figsize}{0.92}
\begin{document}
\title{Constraints on the Topology of the Universe: Extension to General Geometries} 

\date{\today} 

\author{Pascal M.~Vaudrevange} 
\affiliation{DESY, Notkestrasse 85, 22607 Hamburg, Germany}
\affiliation{CERCA \& Department of Physics, Case Western Reserve University, 10900 Euclid Ave, Cleveland, OH 44106, USA} 

\author{Glenn D.~Starkman}
\affiliation{CERCA, ISO \& Department of Physics, Case Western Reserve University, 10900 Euclid Ave, Cleveland, OH 44106, USA} 

\author{Neil J.~Cornish}
\affiliation{Department of Physics, Montana State University, Bozeman, Montana 59717, USA} 

\author{David N.~Spergel}
\affiliation{Department of Astrophysical Sciences, Princeton University, Princeton, New Jersey 08544, USA}

\preprint{DESY 12-090}
\begin{abstract}
  We present an update to the search for a non-trivial topology of the
  universe by searching for matching circle pairs in the cosmic
  microwave background \cite{Cornish:2003db} using the WMAP 7 year
  data release. We extend the exisiting bounds to encompass a wider
  range of possible topologies by searching for matching circle pairs
  with opening angles $10^\circ\le\alpha\le 90^\circ$ and separation
  angles $11^\circ\le\theta\le180^\circ$. The extended search reveal
  two small anomalous regions in the CMB sky.  Numerous pairs of
  well-matched circles are found where both circles pass through one
  or the other of those regions.  As this is not the signature of any
  known manifold, but is a likely consequence of contamination in
  those sky regions, we repeat the search excluding circle pairs where
  both pass through either of the two regions.  We then find no
  statistically significant pairs of matched circles, and so no hints
  of a non-trivial topology.  The absence of matched circles increases
  the lower limit on the length of the shortest closed null geodesic
  that self-intersects at our location in the universe (equivalently
  the injectivity radius at our location) to $98.5\%$ of the diameter
  of the last scattering surface or approximately $26$Gpc.  It extends
  the limit to any manifolds in which the intersecting arcs of said
  geodesic form an angle greater than $10^\circ$.
  \end{abstract}
\maketitle
\section{Introduction}
The search for a non-trivial topology of the universe has enjoyed a
long and fascinating history. Using different methods -- from
searching for specific topologies to the more general circles in the
sky approach \cite{Cornish:1997ab} -- the cosmic microwave background
(CMB) has been analyzed extensively, looking for any signs that light
from the same object reaches us by more than one path (see
\cite{Levin:2001fg} for a review of the various suggested methods,
including
\cite{Cornish:1997hz,Cornish:1997ab,Cornish:1997rp,Cornish:2003db,ShapiroKey:2006hm,
  Bond:1996qs, Bond:1997ym, Souradeep:1998yk, Bond:1999te,
  Bond:1999tf, LachiezeRey:1995kj, Lehoucq:1996qe, Luminet:1999qh,
  Uzan:1998hk, Uzan:1999de, Lehoucq:2000hf,
  Riazuelo:2003ud,Weeks:2003xq,Riazuelo:2006tb,Niarchou:2006be,Niarchou:2007nn}
). So far, all specialized efforts to detect specific topologies as
well as the search for matching opposing circles in the sky have
failed to detect any sign of a non-trivial topology of our universe.

\begin{figure}
  \includegraphics[width=\linewidth]{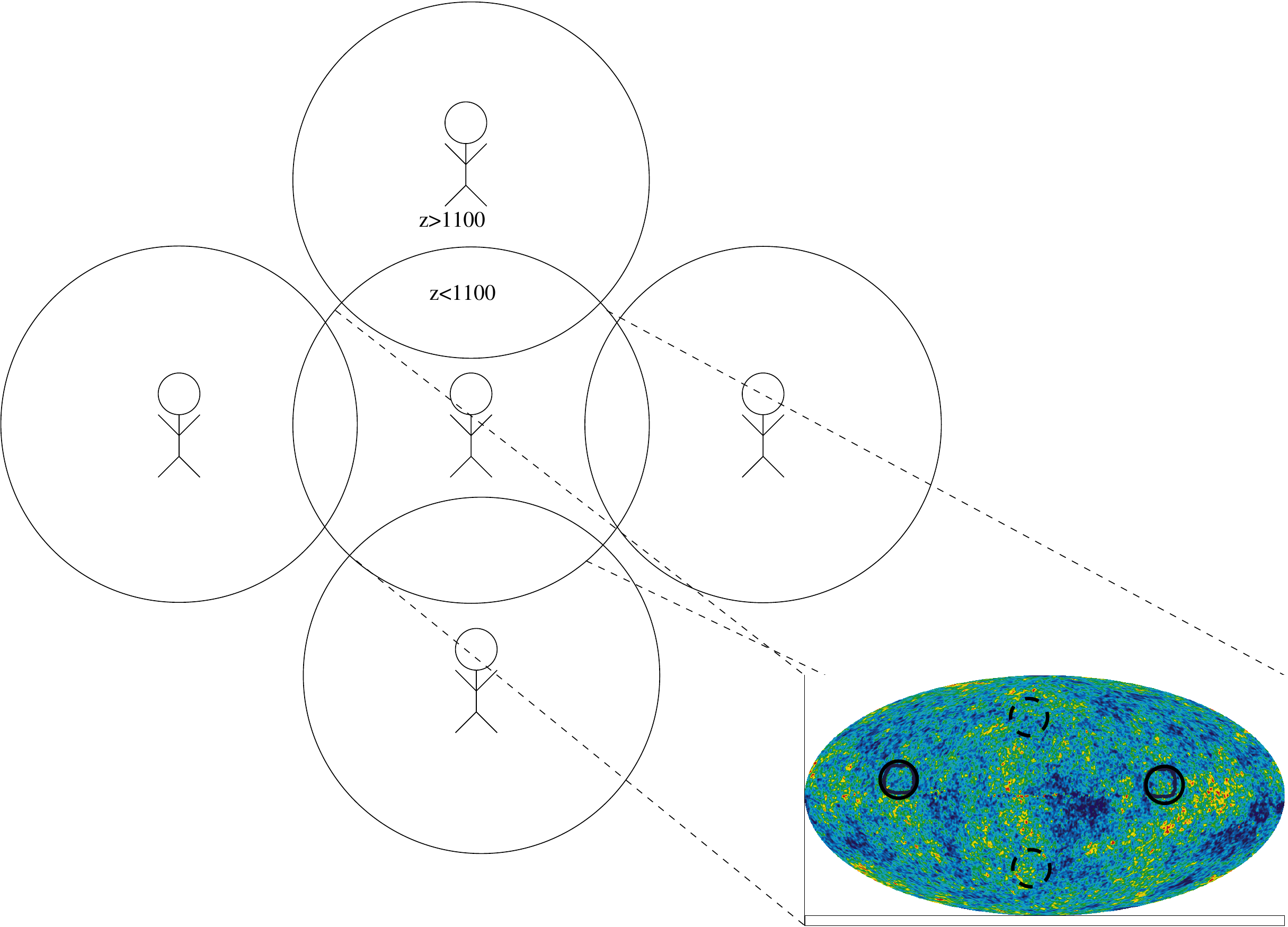}
  \caption{Schematic geometry of the circles-in-the-sky method. If the
    topology of the universe is a torus, the observer, located at the
    center of the 2-sphere of the surface of last scattering, sees
    matching circles of temperature fluctuations on opposite sides of
    the CMB sky.}
  \label{fig:schematic_setup}
\end{figure}

The circles-in-the-sky method, which we adopt in this paper, is based
on the following intuitive picture. For illustrative purposes, assume
that the true topology of the universe is a 3-torus, with unit cell
size smaller than the Hubble horizon (see
Figure~\ref{fig:schematic_setup}). This can be thought of as a tiling
of flat space by identical cubes.  An observer, such as ourselves,
performing a series of cosmic microwave background (CMB) observations
somewhere in one of the cubes, has clones identically located in each
of the other cubes performing the identical series of observations.
Centered around the observer is the 2-sphere of the surface of last
scattering at $z\approx 1100$, with CMB fluctuations imprinted on
it. Around the clone of the observer on the right, there is another
2-sphere of the surface of last scattering. The intersection of both
2-spheres is given by a circle. Both observers will look at the same
ring of temperature fluctuations -- albeit from different
``sides''. Both observers are in fact identical, so an observer will
see a matching pair of circles: one to the left and to the
right. Hence, comparing temperature fluctuations along circles
potentially yields information about the topology of our
universe. Going away from toroidal geometries, it becomes immediately
clear that the separation angle $\theta$ (the angle between the
centers of the pair of matching circles) need not be $180^\circ$.
Also, depending on the orientability or non-orientability of the
manifold, circle pairs might have matching temperature fluctuations
either both going clockwise around the circles (non-orientable) or one
going clockwise and the other anti-clockwise (orientable).

So far, the search for matching anti-podal circles, i.e. circles with
separation angles $180^\circ$, or nearly anti-podal circles
\cite{Cornish:2003db}, like other topology searches, has only yielded
lower limits on the size of the Universe, and then only for ``nearly
flat'' topologies.

In this work, we apply the circles-in-the-sky statistics to searches
for circles pairs of all opening angles
$10^\circ\le\alpha\le90^\circ$, and integer separation angles
$11^\circ\le\theta\le 180^\circ$ with both orientations.  This extends
the previous searches~\cite{Cornish:2003db,ShapiroKey:2006hm} to cover
almost all possible topologies. We find what seems to be a systematic
effect at two special positions in the sky that produces spurious
signals for osculating circles.  Otherwise, we see no evidence of
non-trivial topology.

\section{Methodology}

As already outlined in the introduction, the most intuitive way to
search for a non-trivial topology of the universe is by looking for
matching pairs of circular temperature fluctuation patterns in the
CMB.

In order to determine the underlying topology of the universe, one
would need to scan the full CMB map -- at a Healpix\footnote{{\tt
    http://healpix.jpl.nasa.gov}} resolution of $N_{\text{side}}=512$
corresponding to $3\times10^6$ pixels -- for matching circle pairs. To
conduct a search for all topologies, i.e. over all opening angles of
the circles and all possible separation angles, an enormous number of
circle pairs would need to be analyzed. Previously reported searches
focussed on nearly flat geometries, where matching circle pairs are at
almost opposite positions on the sky, {\it i.e.} $\sim 180^\circ$
apart (see \cite{Cornish:2003db, ShapiroKey:2006hm}).

Thanks to increasing computing power, we can now take a more general
approach. In order to obtain an acceptable time frame for completing
this project, we superimposed a search grid of $N_{\text{side}}=128$
onto the map, resulting in $2\times10^{10}$ circle pairs that need to
be compared. We compute the goodness of the match for a given circle
pair on the full resolution, $N_{\text{side}}=512$, map. Using a
resolution of $512$ pixels along a given circle costs about $10^4$
operations per compared pair, leading to a total number of operations
of $2\times10^{14}$ per opening angle. On a $3$GHz CPU, this takes
about $20$ hours for a single opening angle. Scanning over $200$
opening angles corresponds to $4000$ CPU hours, easily feasible on
modern computer clusters.

Thus we search for all possible topologies, albeit on a somewhat lower
resolution grid. Note that we need to run the search twice, once for
orientable manifolds and once for non-orientable manifolds, {\it i.e.}
once for circle pairs that are oppositely oriented, and once for
circle pairs that are oriented in the same way.

\begin{figure}
  \includegraphics[width=.4\textwidth]{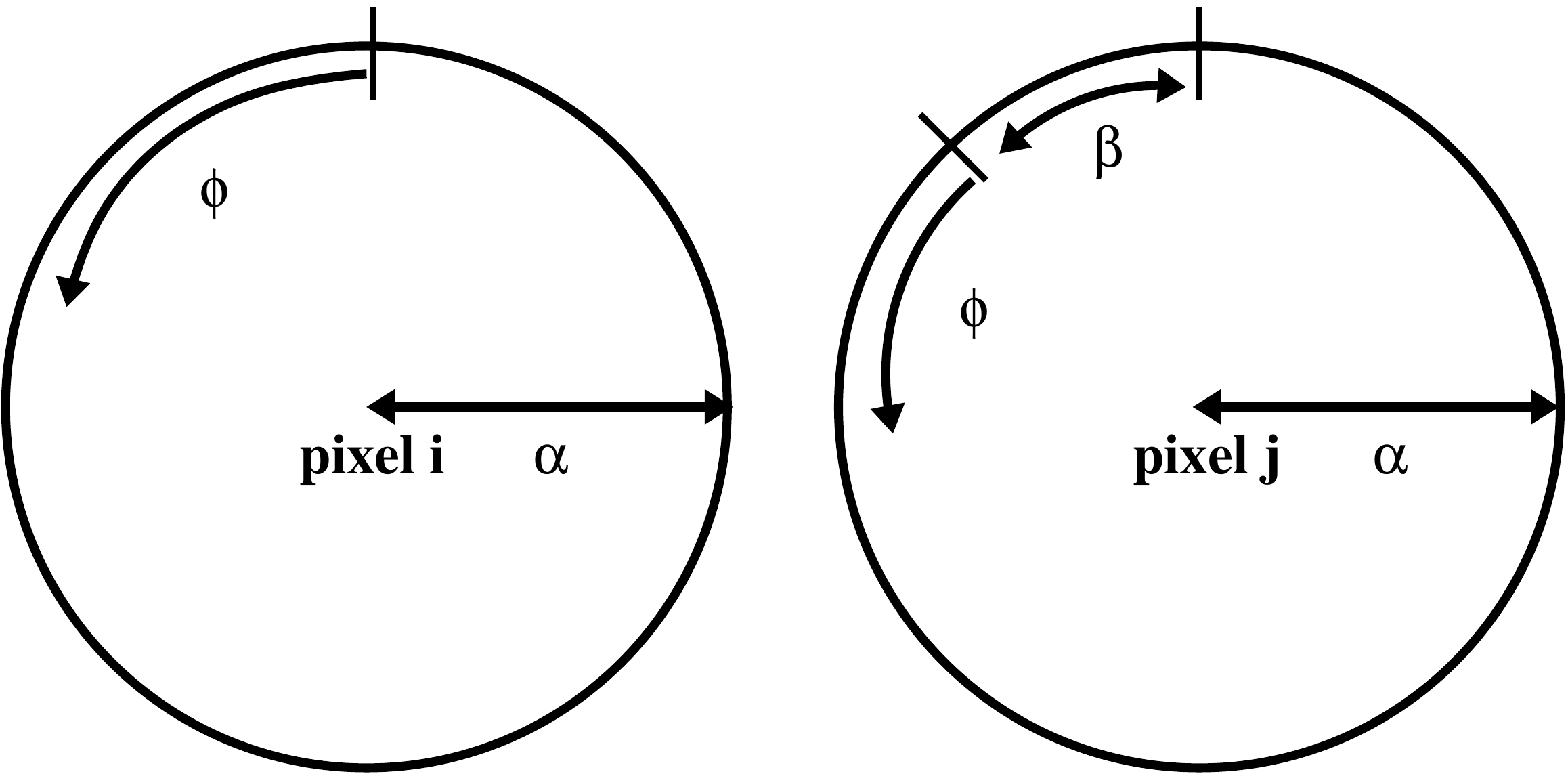}
  \caption{A circle of opening angle $\alpha$ centered at pixel $i$,
    and a circle centered at pixel $i$ with its phase shifted by
    $\beta$.}
  \label{fig:circle_geometry}
\end{figure}

\begin{figure}
  \includegraphics[width=0.3\textwidth]{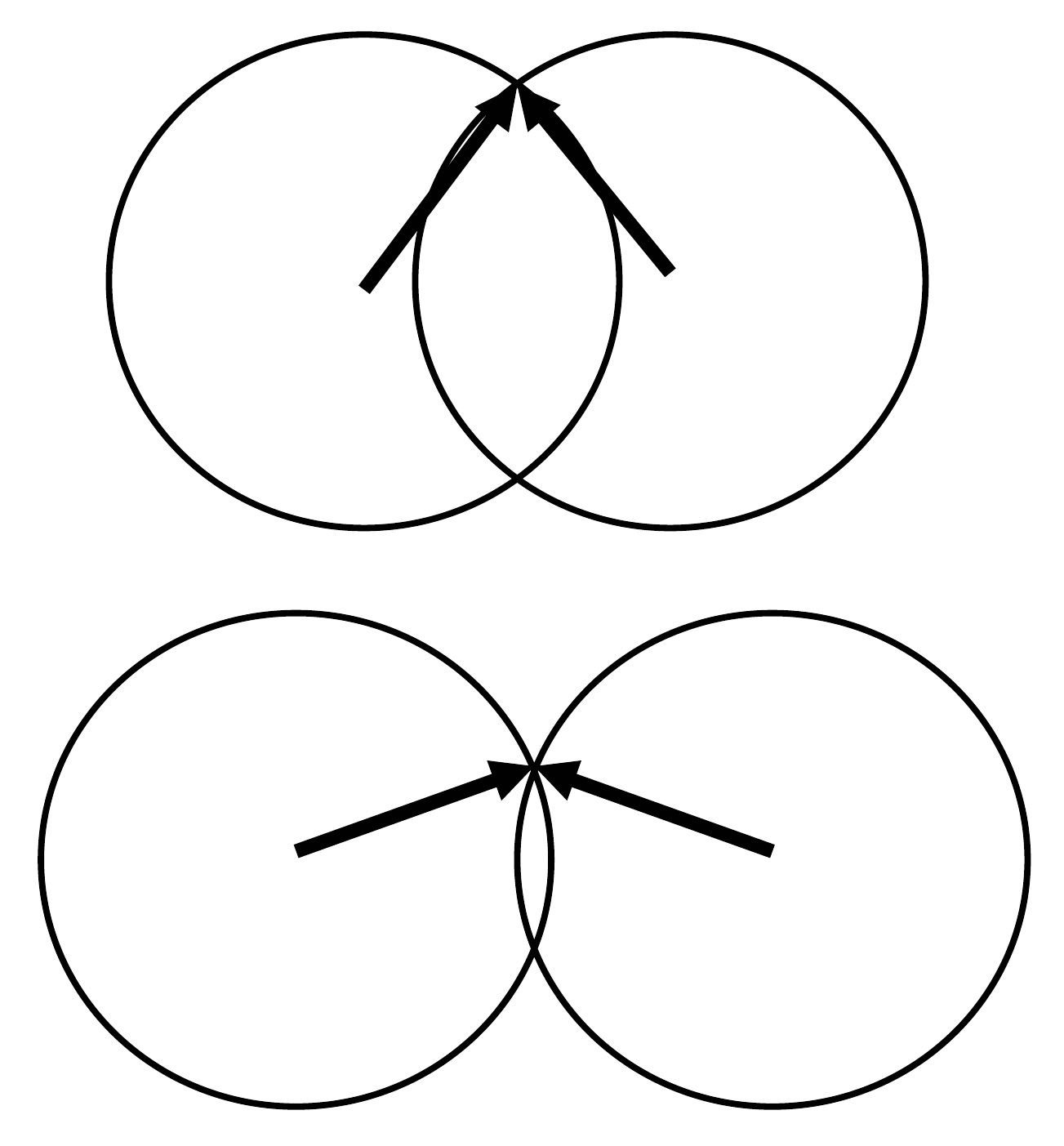}
  \caption{Doppler contribution to the matching circle signal. Small
    circles (bottom) have opposite Doppler contributions, whereas
    large circles (top) obtain similar Doppler contributions. Thus
    large circles show a smaller deviation from $\Smax=1$.}
  \label{fig:doppler}
\end{figure}

A first Ansatz for a circle statistic $S_{ij}$ to measure the match
between circles $i,j$ would be the convolution of the temperature
fluctuations along two circles. As discussed in
\cite{ShapiroKey:2006hm}, this would lead to a dominance of long
wavelength (small $m$) Fourier modes along the circles, making
$S_{ij}$ rather insensitive to small-scale fluctuations. To compensate
for this, \cite{Cornish:2003db} introduced an additional factor of the
wavenumber $m$ in the convolution (this is equivalent to the usual
factor of $\ell(\ell+1)$ used to scale the two dimensional power
spectrum). Thus, in order to compare two circles of opening angle
$\alpha$ centered around pixel numbers $i$ and $j$, we employ the
circle statistic
\begin{eqnarray}\label{eq:def:smax}
  S_{ij}(\alpha, \beta)&=&\frac{1}{\sum_{m=0}^{n/2} m
    \left(|T_{im}(\alpha)|^2 +|T_{jm}(\alpha)|^2\right)}\,\nonumber\\ 
  &\times&\left(\sum_{m=0}^{n/2}
  mT_{im}^*(\alpha) T_{jm}(\alpha) e^{-\frac{2\pi i
      m\beta}{n}}\right.\nonumber\\
  &&\left.+\sum_{m=0}^{n/2} (n-m) T_{im}^*(\alpha)
  T_{jm}(\alpha) e^{-\frac{2\pi i m\beta}{n}}\right),\quad
\end{eqnarray}
where $\beta$ is the relative phase between the two circles and
$T_{im}(\alpha)$ is the Fourier transform of the temperature
fluctuation $\frac{\Delta T}{T}$ around circle $i$
\begin{eqnarray}
  \frac{\Delta T}{T}(i,\phi)=\sum_{m=0}^{n-1}T_{im} e^{\frac{2\pi im\phi}{n}}\, ,
\end{eqnarray}
where $n=512$ along the circle (see Figure~\ref{fig:circle_geometry}).
For a search for orientable manifolds, we replace all occurrences of
$T_{im}^*(\alpha)$ in the numerator by $T_{im}(\alpha)$. Notice that
we use the conventions of the {\tt libfftw}
package\cite{fftw}\footnote{{\tt http://www.fftw.org}}: positive
frequencies are stored in the first half of the array $T_{im},
m=0\dots \frac{n}{2}$, and negative frequencies are stored in
backwards order from the end of the array, making
$T_{im}=T^*_{i(n-m)}$\footnote{Using the real--to--complex and
  complex--to--real routines enables us to save the Fourier transforms
  for a single circle of a given opening angle in an complex array of
  length $\frac{n}{2}$ instead of $n$.}. We then use as statistic the
maximum of $S_{ik}(\alpha, \beta)$ over all relative phases $\beta$
\begin{eqnarray}
  \Smax(\alpha)&\equiv&\max_\beta S_{ij}(\alpha, \beta)\,.
\end{eqnarray}
For perfectly matching circles, $\Smax(\alpha)=1$, and for perfectly
uncorrelated circles, $\Smax(\alpha)\approx 0$. In practice, these
ideal values are not realized due to noise contribution from several
different effects. First of all, the Doppler effect at the surface of
last scattering creates a different signal depending on the position
of the observer, making larger matching circles closer to
$\Smax\approx 1$ whereas smaller circles will have $\Smax<1$ ( see
Figure~\ref{fig:doppler}).
Another contribution comes from line--of--sight effects, especially
the ISW effect, as the CMB photons from different directions traverse
through different patches of space. The combination of these effects
reduces the signal in $\Smax$, making potential matching circles less
than perfect (see Figure~1 in \cite{Cornish:2003db}).

For this analysis, we used the WMAP7 temperature maps
\cite{Jarosik:2010iu}.  Outside the WMAP
Kp12\footnote{http://lambda.gsfc.nasa.gov/product/map/dr1/intensity\_mask.cfm}
sky mask, we used the same template cleaning method as
\cite{Gold:2010fm}.  Inside the Kp12 sky cut, we used the WMAP ILC
map\footnote{http://lambda.gsfc.nasa.gov/product/map/dr4/ilc\_map\_get.cfm}.
While this choice implied that the noise properties and resolution of
the map differed between the two regions, the effect on the circle
search was relatively small as the ILC map was only used for 5.8\% of
the pixels.

In order to estimate the significance of potential spikes in the
statistics, we created approximate random realizations of the
CMB\footnote{In contrast to the WMAP 1 year data that was examined in
  \cite{Cornish:2003db}, here the statistics is not
  noise-dominated. In particular this means that the estimator for the
  false detection rate devised in \cite{Cornish:2003db} is not
  applicable for the WMAP7 year data.}. To this end, we compute the
$a_{\ell m}$'s of the cleaned CMB map. Then, we scramble them by
randomly interchanging the $m$ index for fixed $\ell$'s, and compute a
``random'' CMB map from the scrambled $a_{\ell m}$'s. Using this map,
we compute the statistics for matching (non-)orientable circle pairs
for opening angles $\alpha=20^\circ, 50^\circ, 80^\circ$ and all
integer separation angles $\theta$. Repeating this $1000$ times, we
obtain an estimate of the probability density function (pdf) of
$S^{\text{max}}$. While there are enough samples to give a reliable
estimate of the $95.4\%$ ($2\sigma$) confidence level (CL), the quoted
$99.7\%$ ($3\sigma$) CL is at most a rough estimate.  If the sky noise
were isotropic, then this randomization process would generate
simulated maps with the same statistical properties as the WMAP
observations.  However, because of anisotropies due to spatial
variations in the WMAP noise (which is larger near the ecliptic plane)
and variations in the resolution of our map (due to the need to use
the ILC map in the galactic plane), these simulated maps only
approximate the WMAP sky maps.

\section{Results}

Implementing the procedure outlined in the previous section, we
present the results of the searches and describe the systematic
effects we encountered, both for orientable and non-orientable
topologies.

Apart from a possible signal, there are simple random statistical
fluctuations which are expected to exceed the $95\%$ CL. To estimate
the number of random fluctuations, we note that this can be viewed as
a series of Bernoulli trials with probability $p=0.05$ for success,
i.e. for a spike above $95\%$ CL. Per separation angle $\theta$, we
probed $n=190$ values of the opening angle $\alpha$ (by choosing it to
lie on a grid deriving from the position of the rings in the Healpix
scheme for $n_{\text{side}}=128$. The probability distribution
function for having $i$ excursions above the $2\sigma/ 95.4\%$ CL for
a sequence of $n$ Bernoulli trials is given by the Binomial
distribution
\begin{eqnarray}
   p^i (1-p)^{n-i} \left(\begin{array}{c}n\\i\end{array}\right)\,,
\end{eqnarray}
whose expectation value is given by 
\begin{eqnarray}
  \langle N\rangle=\sum_{i=1}^{n} i p^i (1-p)^{n-i} \left(\begin{array}{c}n\\i\end{array}\right)=np=8.7\,.
\end{eqnarray}
Thus, we expect about $9$ spikes above the $2\sigma$ CL per separation
angle $\theta$. Similarly, we expect about $0.6$ spikes above the
$3\sigma$ CL per separation angle $\theta$.

\subsection{Search for Orientable Topologies}
\begin{figure*}
  \includegraphics[width=\figsize\linewidth]{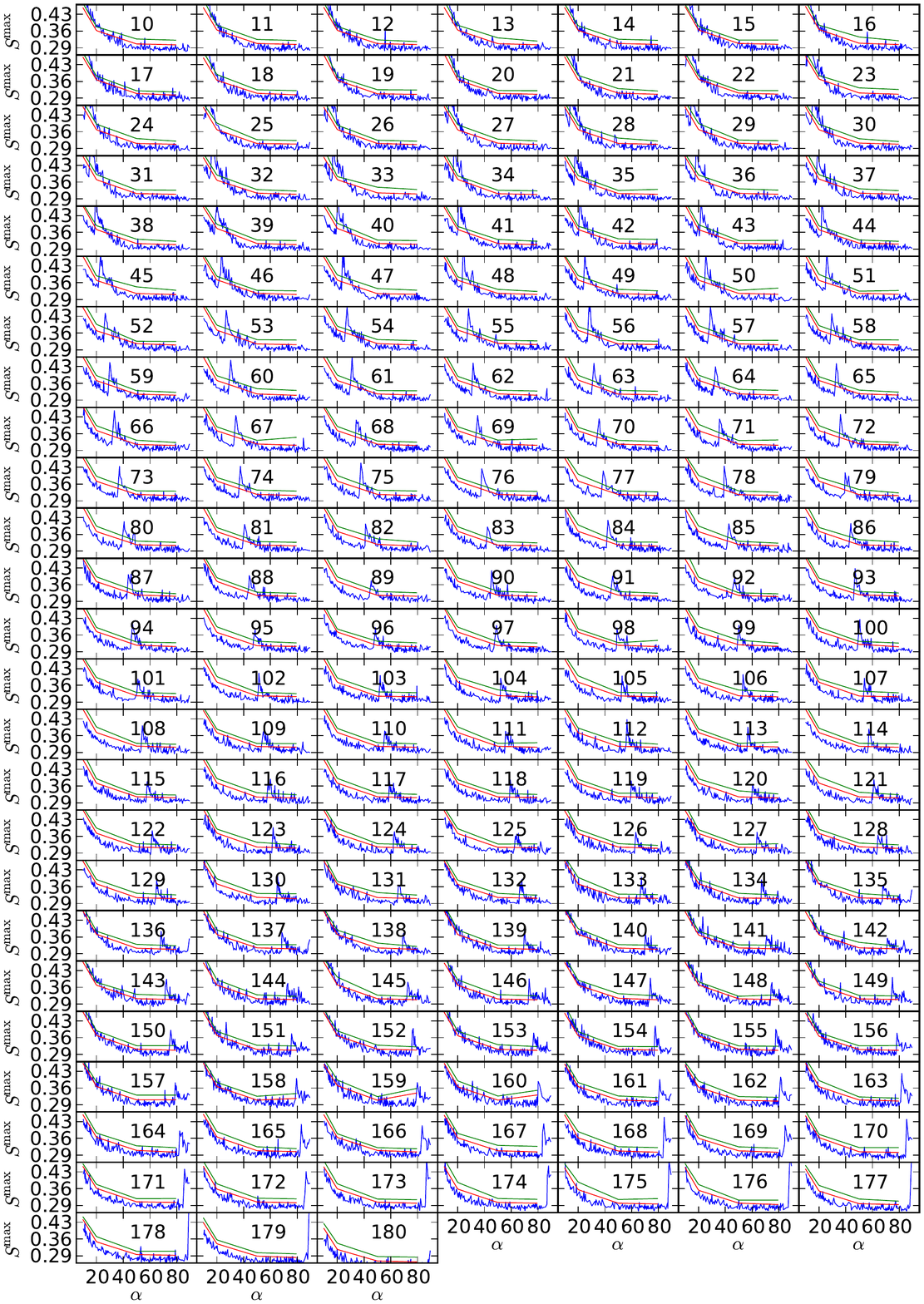}
  \caption{Searching for orientable topologies: The statistics $S$
    (solid blue line) as defined in Equation~\ref{eq:def:smax} as a
    function of opening angle $\alpha$ and separation angle $\theta$
    (indicated by the numbers in the center of each plot). The red/
    green line correspond to the $2\sigma/ 3\sigma$ CL. The spikes
    come from circles osculating at the galactic anti-center and at
    $l=109.44^\circ$, $b=27.8^\circ$ (see
    Figure~\ref{fig:w_orientable_skymap}). Removing circles that touch
    these regions removes all features (see
    Figure~\ref{fig:w_orientable_exclude_galactic_anticenter2_5_exclude_special_position2_5}).}
  \label{fig:w_orientable}
\end{figure*}

Searching for orientable topologies, we find spikes in the
distribution of the statistics $S$ as a function of opening angle
$\alpha$ and separation angle $\theta$ (see
Figure~\ref{fig:w_orientable}).
The red (green) line indicate the $2\sigma (3\sigma)$ CL as determined
from scrambling the $a_{\ell m}$'s which we described above.

\begin{figure}
  \includegraphics[width=\linewidth]{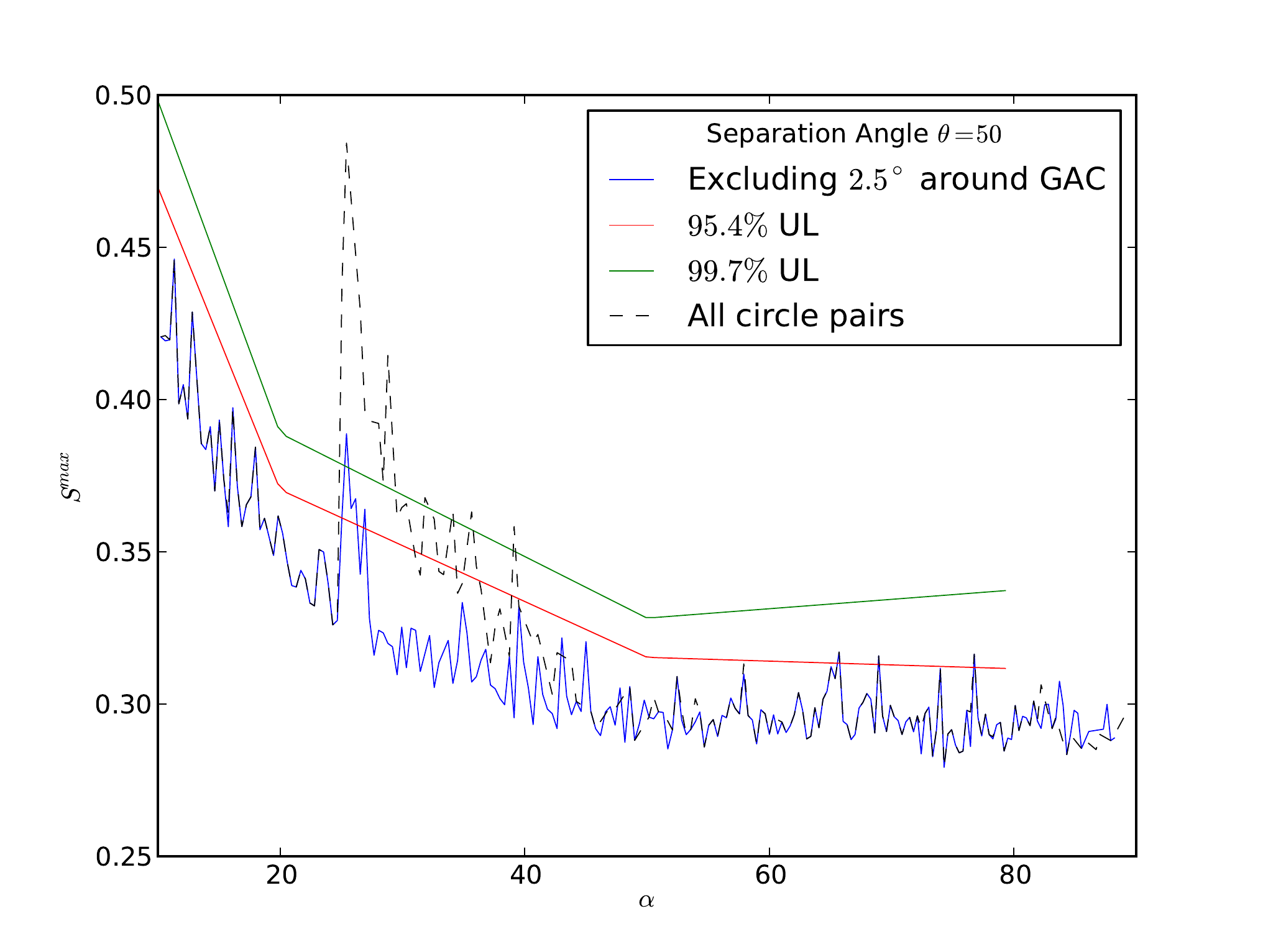}
  \caption{Searching for orientable topologies: For separation angle
    $\theta=50^\circ$, a plot of the statistics $S$ as a function of
    opening angle $\alpha$. Without cutting anything, the dashed black
    line shows an extended feature at
    $\alpha=\frac12\theta=25^\circ$. Ignoring all circle pairs that
    overlap within $2.5^\circ$ degrees of the galactic anti-center
    (GAC), the feature mostly disappears (solid blue line). The red
    (green) solid lines show the $95.4\%$ ($99.7\%$) CL on $S$,
    obtained by scrambling the $a_{\ell m}$ (see the main text). See
    Figure~\ref{fig:w_orientable_skymap} for an explanation of the
    systematic spikes in $S$ as well as
    Figure~\ref{fig:w_orientable_exclude_galactic_anticenter2_5_exclude_special_position2_5}
    for the final result after excising the offending circles.}
  \label{fig:bump_in_theta50_orientable}
\end{figure}

\begin{figure}
  \includegraphics[width=\linewidth]{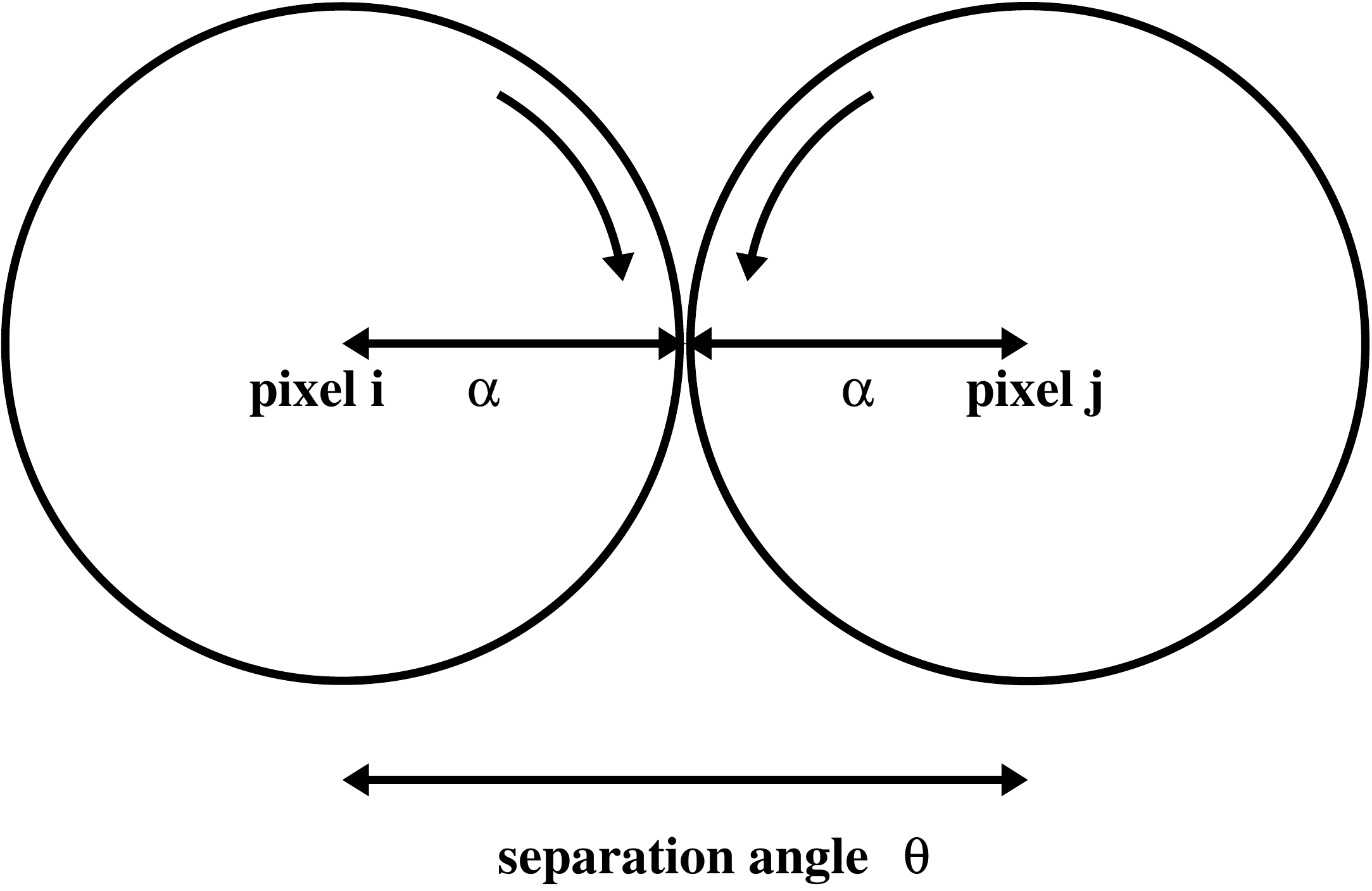}
  \caption{As soon as the separation angle $\theta$ between the two
    circles becomes approximately twice their opening angle $\alpha$,
    $\theta<2\alpha$, the circles start partially overlapping,
    creating matching patterns in the statistics $S^{\text{max}}$ as
    for orientable manifolds, the ``kissing'' segments match.
  }
  \label{fig:bump_from_intersecting_circles}
\end{figure}

\begin{figure*}
  a)\includegraphics[width=0.45\linewidth]{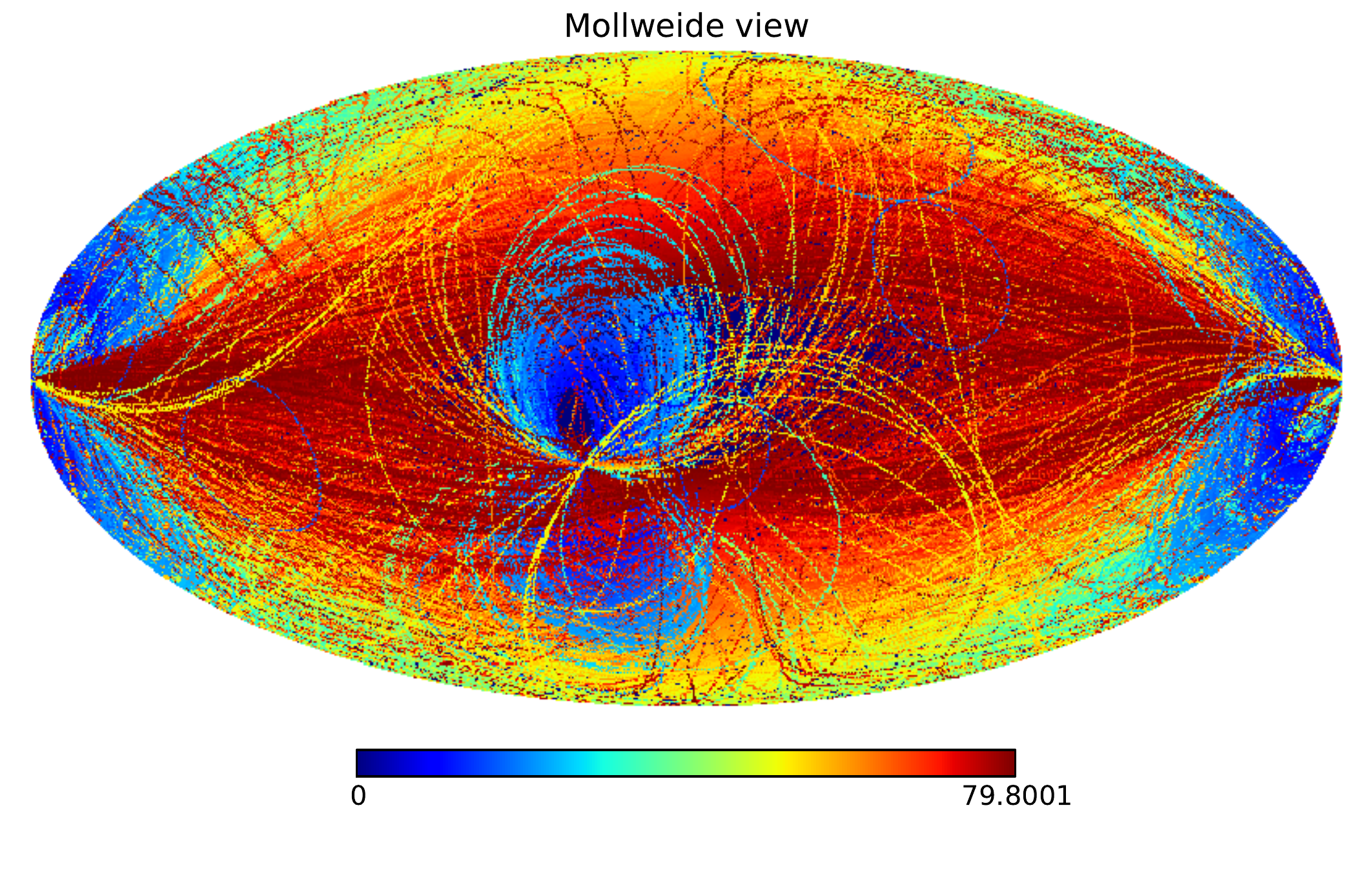}
  b)\includegraphics[width=0.45\linewidth]{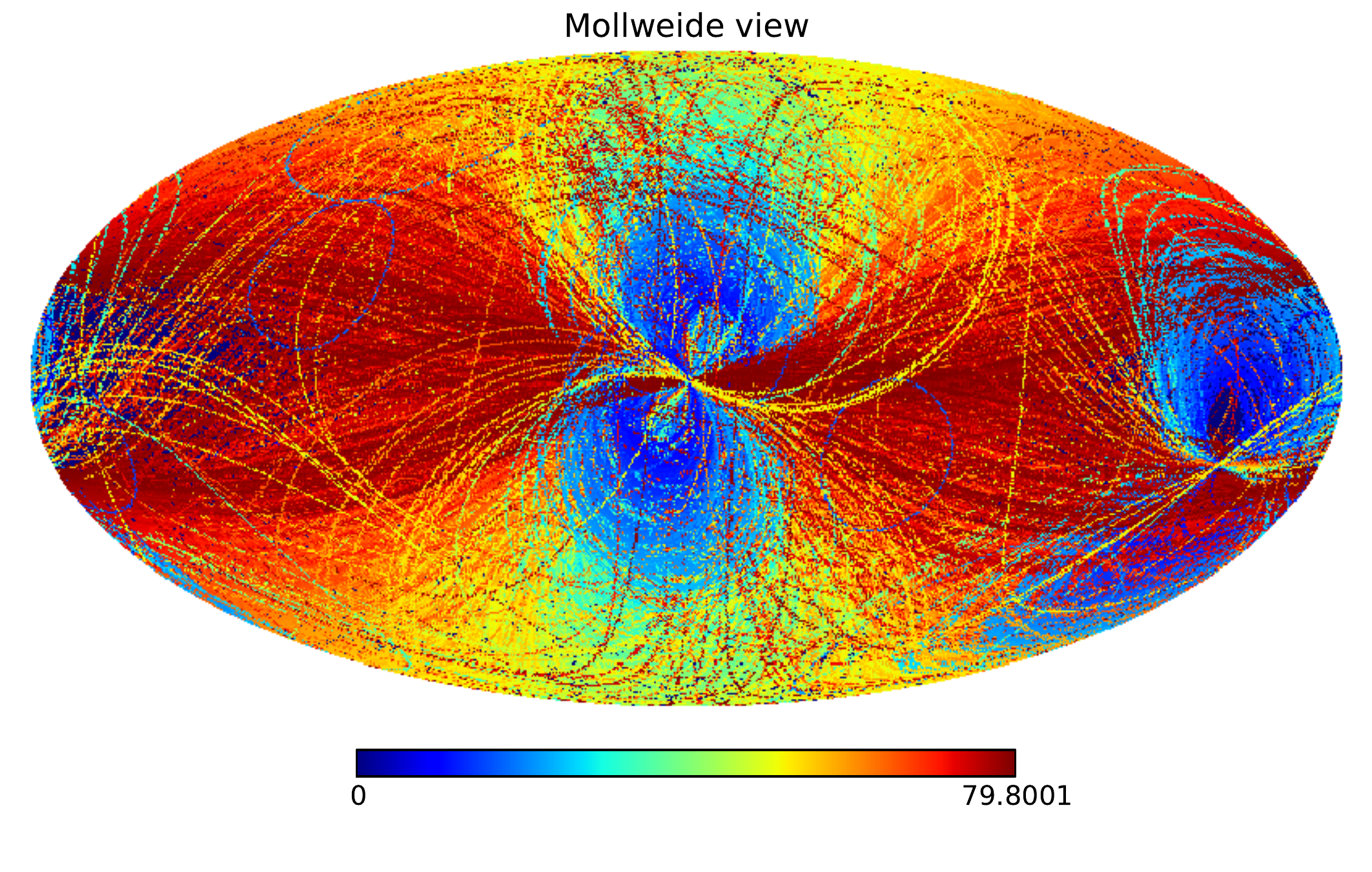}\\
  c)\includegraphics[width=0.45\linewidth]{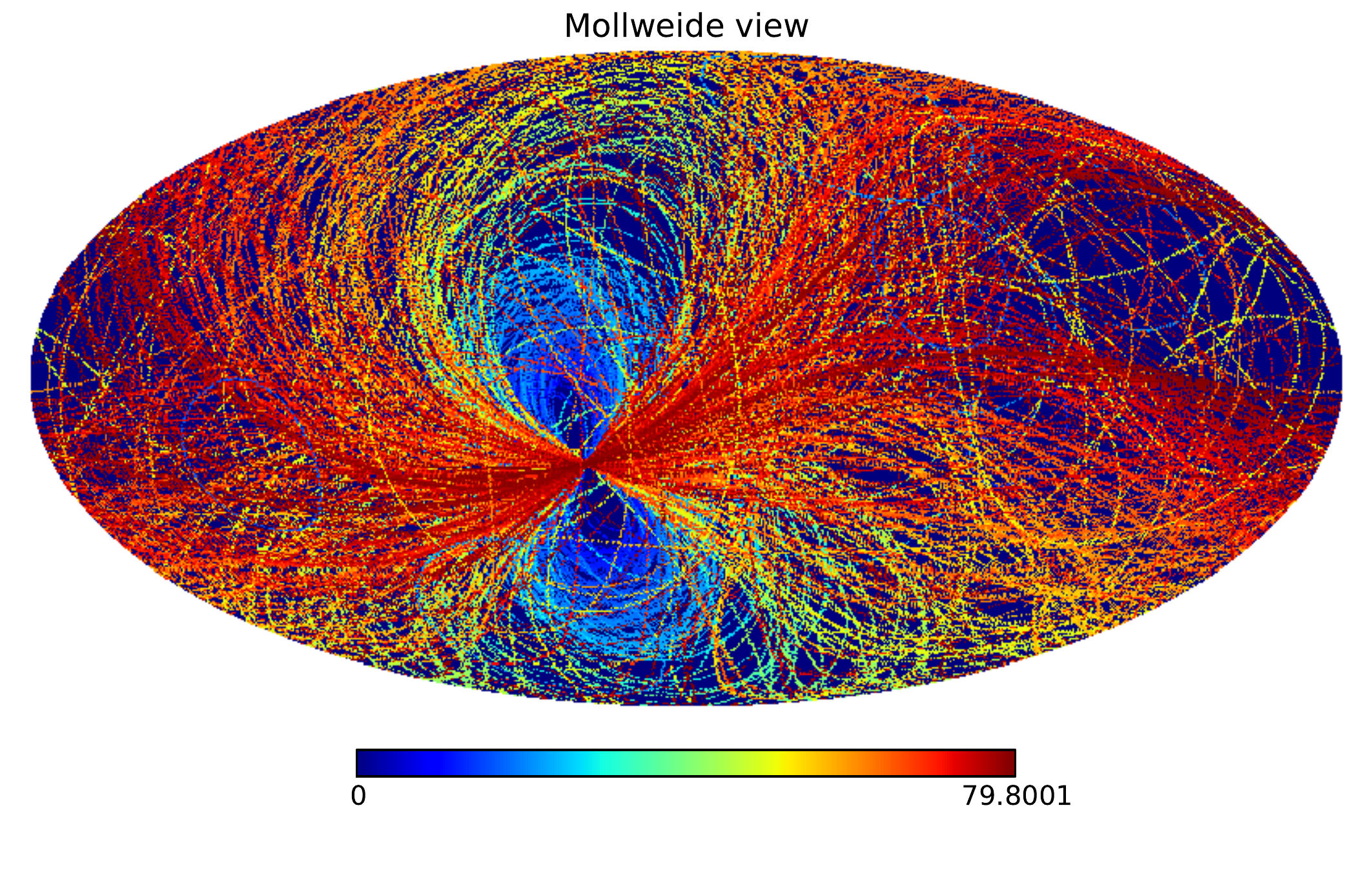}
  d)\includegraphics[width=0.45\linewidth]{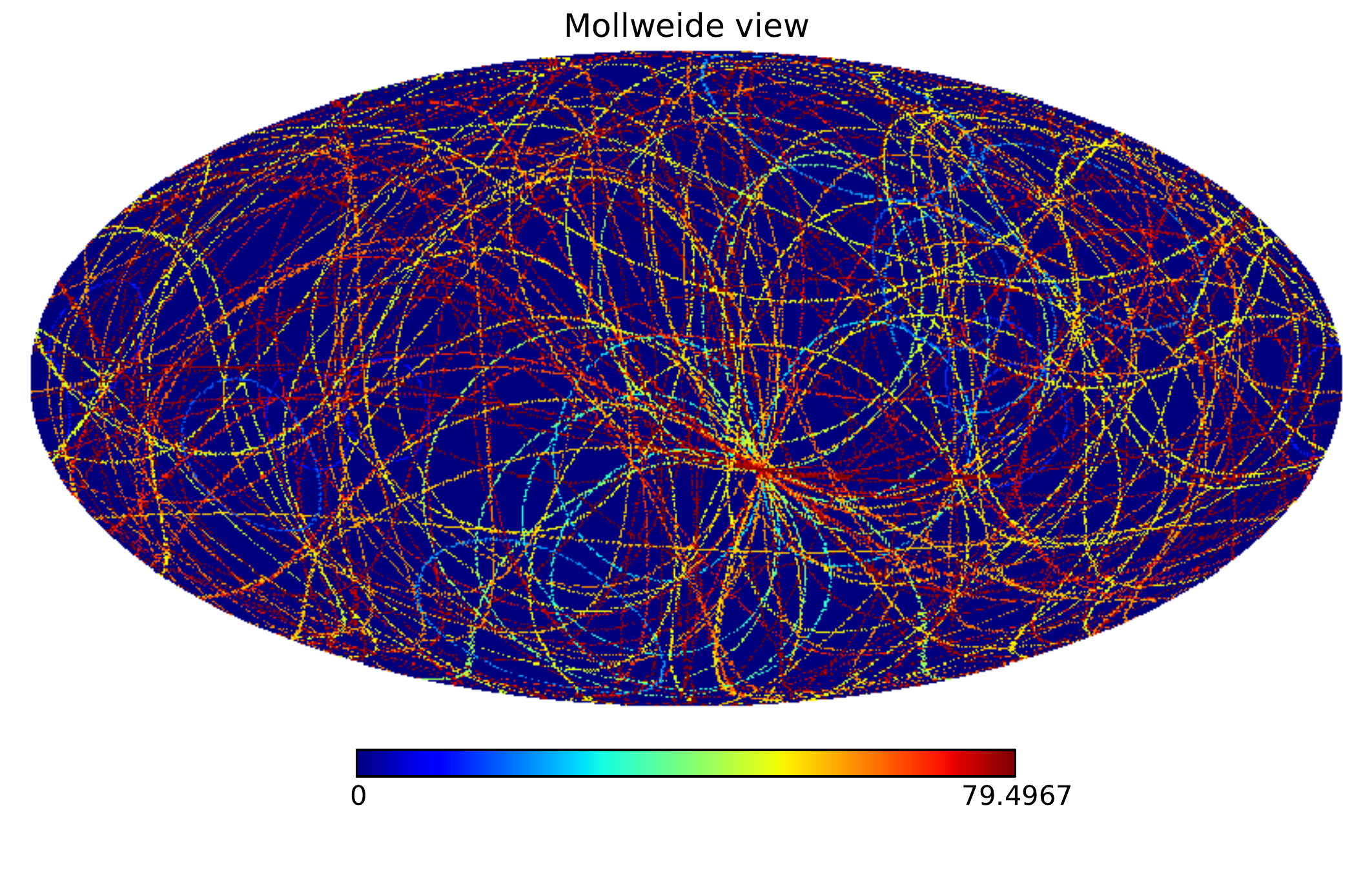}
  \caption{For the search for orientable manifolds: Location of the
    circles pairs with maximal statistics $S$ that lies above the
    $3\sigma$ CL, colored by opening angle $\alpha$. Note that the
    osculating circles at the galactic anti-center and at
    $l=109.44^\circ$, $b=27.8^\circ$ do not hint at a non-trivial
    topology. a) The highest signal comes from circle pairs osculating
    at the galactic anti-center and at $l=109.44^\circ$,
    $b=27.8^\circ$. b) same as a), but rotated by $180^\circ$ such
    that the galactic anti-center is in the middle of the plot. c)
    Removing circle pairs that osculate at the galactic anti-center,
    the highest signal comes from pairs that osculate at
    $l=109.44^\circ$, $b=27.8^\circ$. d) Removing circles that
    osculate either at the galactic anti-center or at
    $l=109.44^\circ$, $b=27.8^\circ$, no special position on the sky
    is apparent.}
  \label{fig:w_orientable_skymap}
\end{figure*}

There is an extended feature whose position (as a function of the
opening angle $\alpha$) is correlated with the separation angle
$\theta$ -- it seems to be located at
$\alpha=\frac12\theta$. Figure~\ref{fig:bump_in_theta50_orientable}
shows the values of $S^{\text{max}}$ as a function of $\alpha$ for two
circles separated by $\theta=50^\circ$, with a bump clearly visible
around $\alpha=25^\circ$.
This can be understood as the effect of two circles partially
overlapping (see Figure~\ref{fig:bump_from_intersecting_circles}). 
If two circles are separated by twice their opening angle $\alpha$
they osculate (``kiss''). For orientable manifolds this means that
their patterns nearly match up along a segment. This effect is almost
independent of the absolute magnitude of the opening angle $\alpha$:
the fraction of the circles that ``kiss'' is independent of $\alpha$
(up to effects of finite pixel size).  However, it is interesting to
note that most of the osculating circles that give the maximal $S$ for
a given separation angle and opening angle osculate at either of two
positions: at the galactic anti-center, or at $l=109.44^\circ$,
$b=27.8^\circ$ (see Figure~\ref{fig:w_orientable_skymap}).

Note that there is no known topology which would lead to such such a
structure of spikes in the $S$ statistics, correlating the opening
angle with the separation angle, confined to just two positions on the
sky and appearing both in the search for orientable topologies as well
as in the search for non-orientable topologies (see next
Subsection). Hence it is justified to remove all circles that touch
either the galactic anti-center or the position $l=109.44^\circ$,
$b=27.8^\circ$.  The existence of this anomaly near the galactic
anti-center is perhaps not surprising.  The anti-center is in the
middle of the ILC portion of the map, and ILC pixels have much more
correlated noise properties than the rest of the map.  The second
point is near the region where the noise properties of the maps are
very non-uniform (see Figure 3 in \cite{Bennett:2003bz}).

First excising circles that osculate at the galactic anti-center,
Figure~\ref{fig:w_orientable_skymap}c), and then those that osculate
at $l=109.44^\circ$, $b=27.8^\circ$,
Figure~\ref{fig:w_orientable_skymap}d), removes most systematic spikes
from the $S$ statistics. The peaks that are left for separation angles
$\theta>170^\circ$ and opening angle $\alpha\approx90^\circ$ are
caused by the fact that circles with $\alpha=90^\circ$ have an angular
diameter of $180^\circ$. Thus, two such circles, when separated by
$\theta\approx180^\circ$ will start overlapping, until for
$\theta=180^\circ$, they coincide, independent of the topology of the
universe.

\begin{figure*}
  \includegraphics[width=\figsize\linewidth]{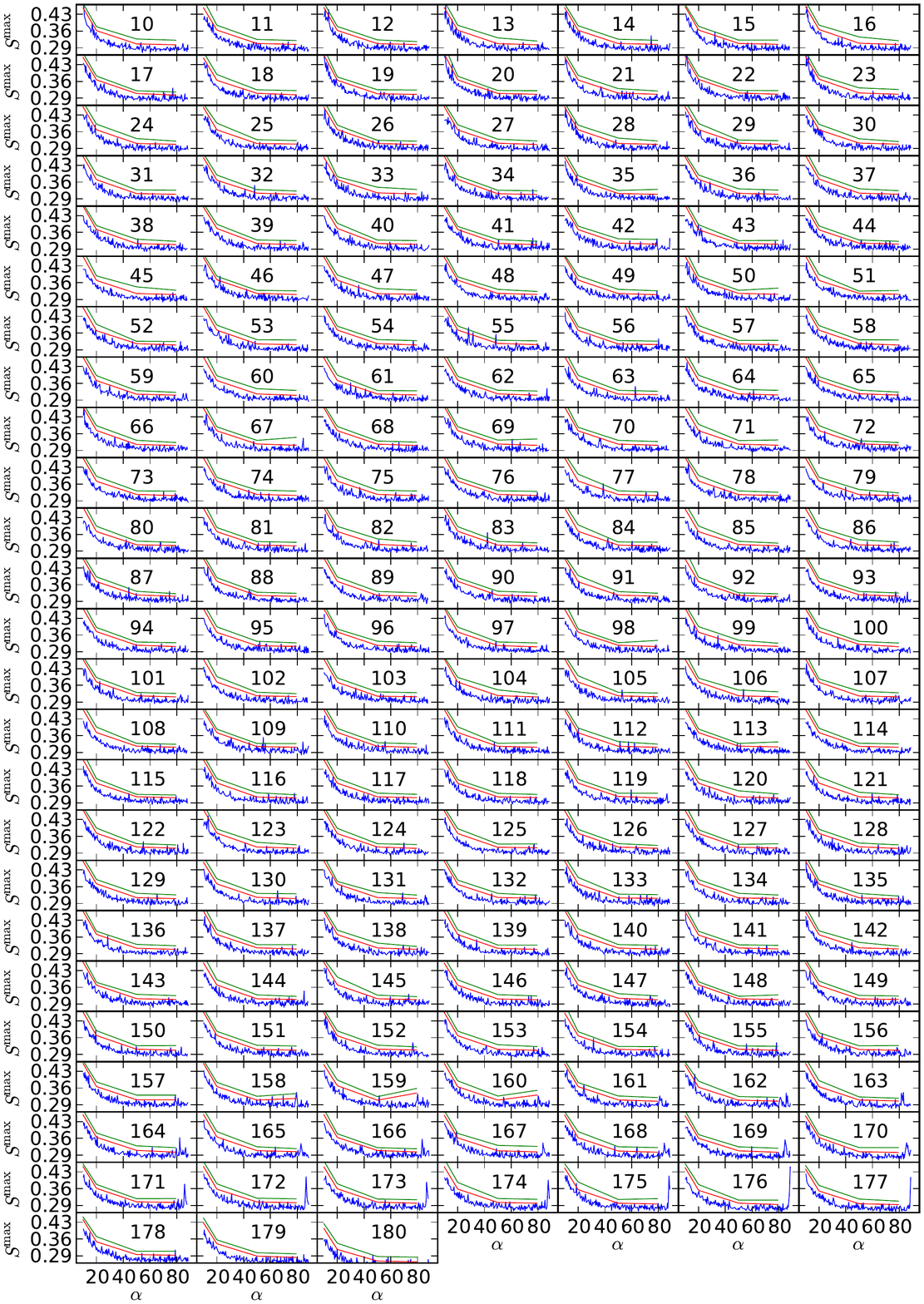}
  \caption{Searching for orientable topologies, final result: The
    statistics $S$ (solid blue line) as defined in
    Equation~\ref{eq:def:smax} as a function of opening angle $\alpha$
    and separation angle $\theta$ (indicated by the numbers in the
    center of each plot). The red/ green line correspond to the
    $2\sigma/ 3\sigma$ CL. We disregard circles that touch either the
    galactic anti-center or $l=109.44^\circ$, $b=27.8^\circ$. The
    peaks for separation angles $\theta>170^\circ$ and opening angle
    $\alpha\approx90^\circ$ are caused by the fact that circles with
    $\alpha=90^\circ$ have an angular diameter of $180^\circ$. Thus,
    two such circles, when separated by $\theta\approx180^\circ$ will
    start overlapping, until for $\theta=180^\circ$, they coincide,
    independent of the topology of the universe. No signs of a
    non-trivial orientable topology are found.}
  \label{fig:w_orientable_exclude_galactic_anticenter2_5_exclude_special_position2_5}
\end{figure*}

Hence
Figure~\ref{fig:w_orientable_exclude_galactic_anticenter2_5_exclude_special_position2_5}
presents the final result of the search for orientable topologies.
The number of spikes above the $2\sigma$ and $3\sigma$ thresholds are
consistent with random fluctuations.  This leads us to conclude that
the data is consistent at 99.7\% CL with the null hypothesis of no
non-trivial, orientable topology.

\subsection{Search for Non-Orientable Topologies}
\begin{figure*}
  \includegraphics[width=\figsize\linewidth]{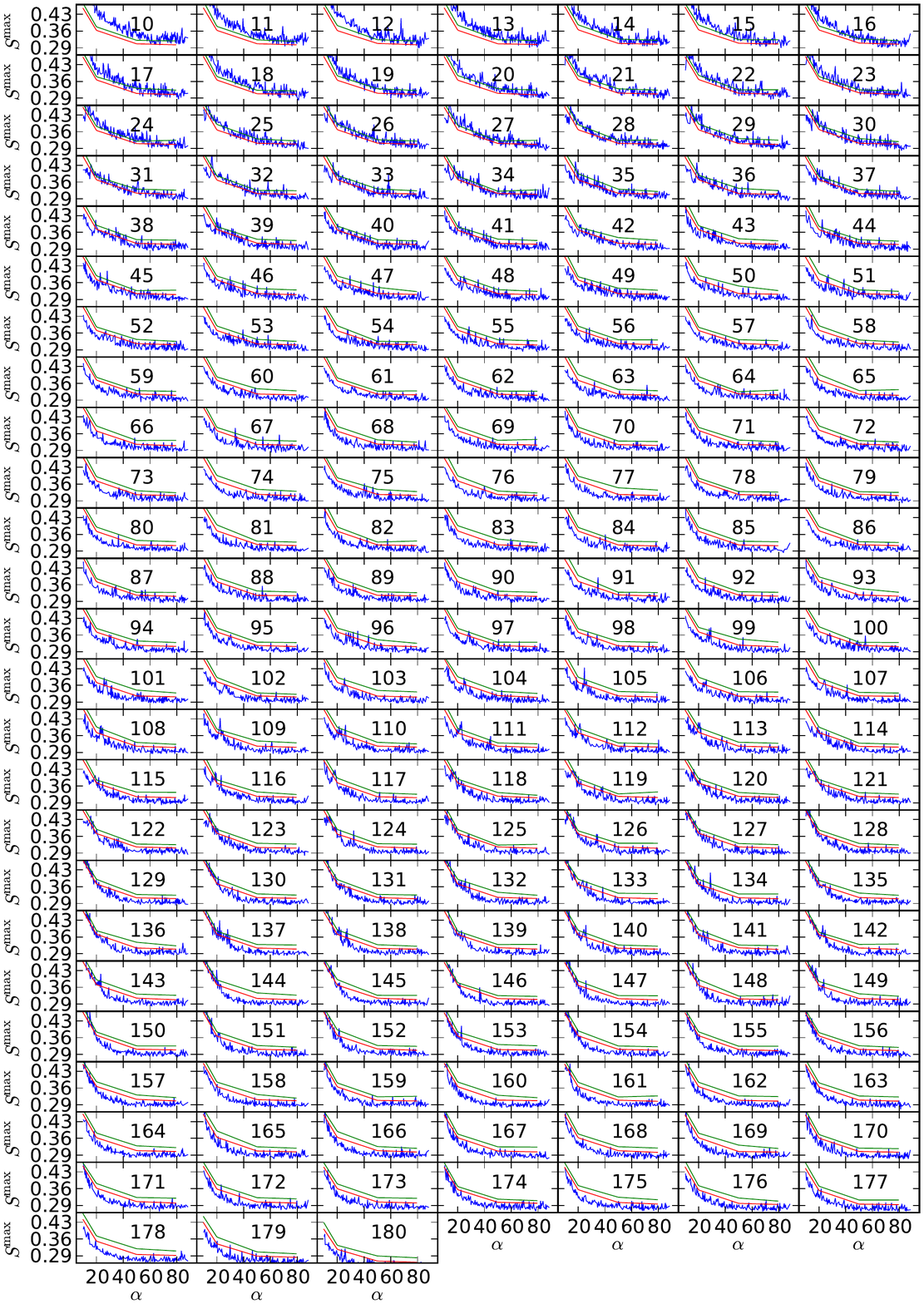}
  \caption{Searching for non-orientable topologies: The statistics $S$
    (solid blue line) as defined in Equation~\ref{eq:def:smax} as a
    function of opening angle $\alpha$ and separation angle $\theta$
    (indicated by the numbers in the center of each plot). The red/
    green line correspond to the $2\sigma/ 3\sigma$ CL. The spikes
    come from circles osculating in the galactic anti-center and at
    $l=109.44^\circ$, $b=27.8^\circ$ (see
    Figure~\ref{fig:w_non_orientable_skymap}). Removing circles that
    touch these regions removes all features (see
    Figure~\ref{fig:w_non_orientable_exclude_galactic_anticenter2_5_exclude_special_position2_5b}).
  }
  \label{fig:w_non_orientable}
\end{figure*}

Performing the search for non-orientable topologies on the cleaned CMB
map, we again find a multitude of excursions above the $3\sigma$ CL
(see Figure~\ref{fig:w_non_orientable}). There, we plot in blue the
statistics $S$ as a function of opening angle $\alpha$ and separation
angle $\theta$ (the former indicated by the numbers in the middle of
each panel). The red (green) line is the $2\sigma (3\sigma)$ CL
obtained by scrambling the $a_{\ell m}$'s as described above. In
particular, for low values of the separation angle $\theta$, the
signal lies consistently above the $3\sigma$ CL.
\begin{figure*}
  a)\includegraphics[width=0.45\linewidth]{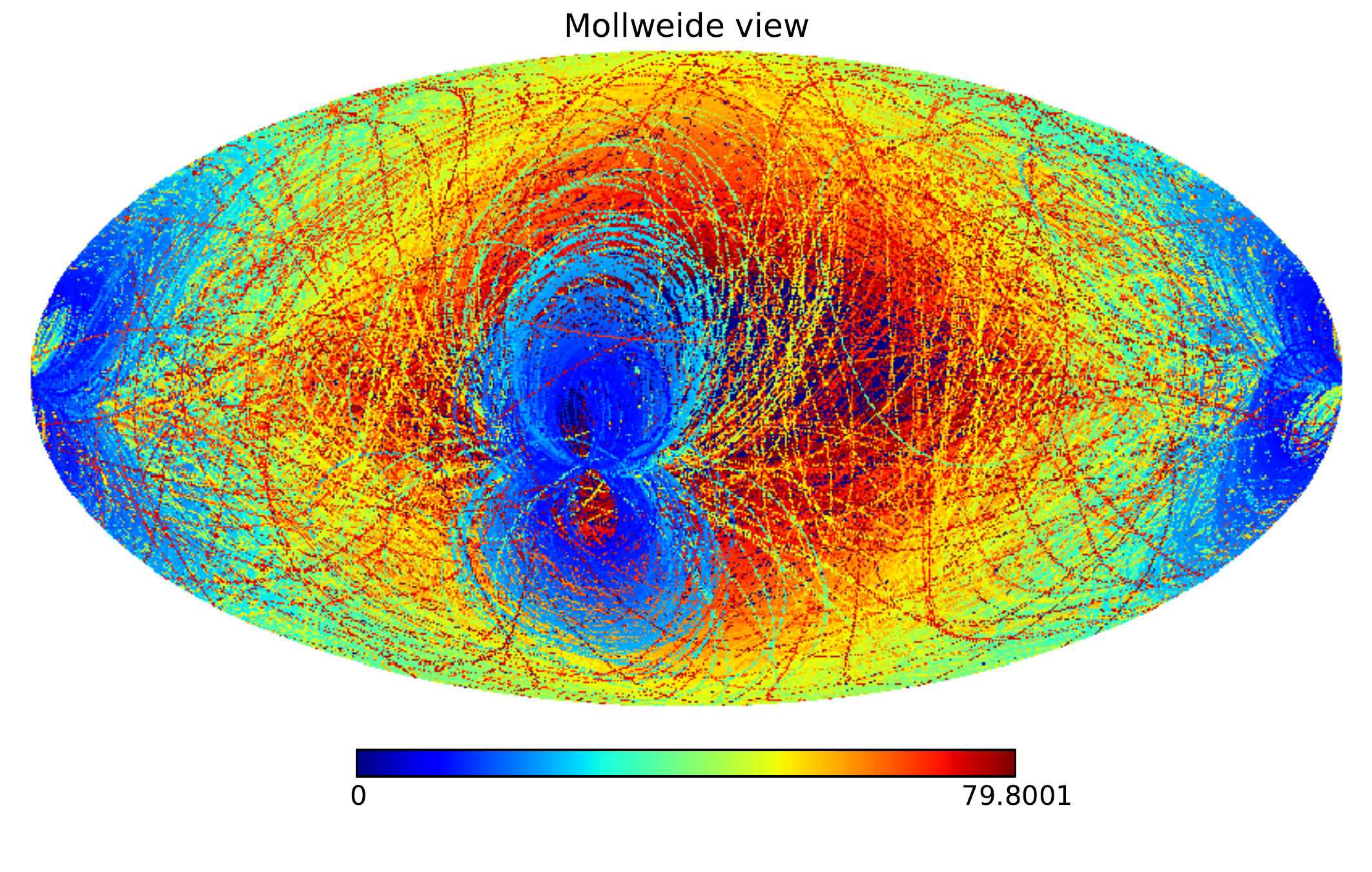}
  b)\includegraphics[width=0.45\linewidth]{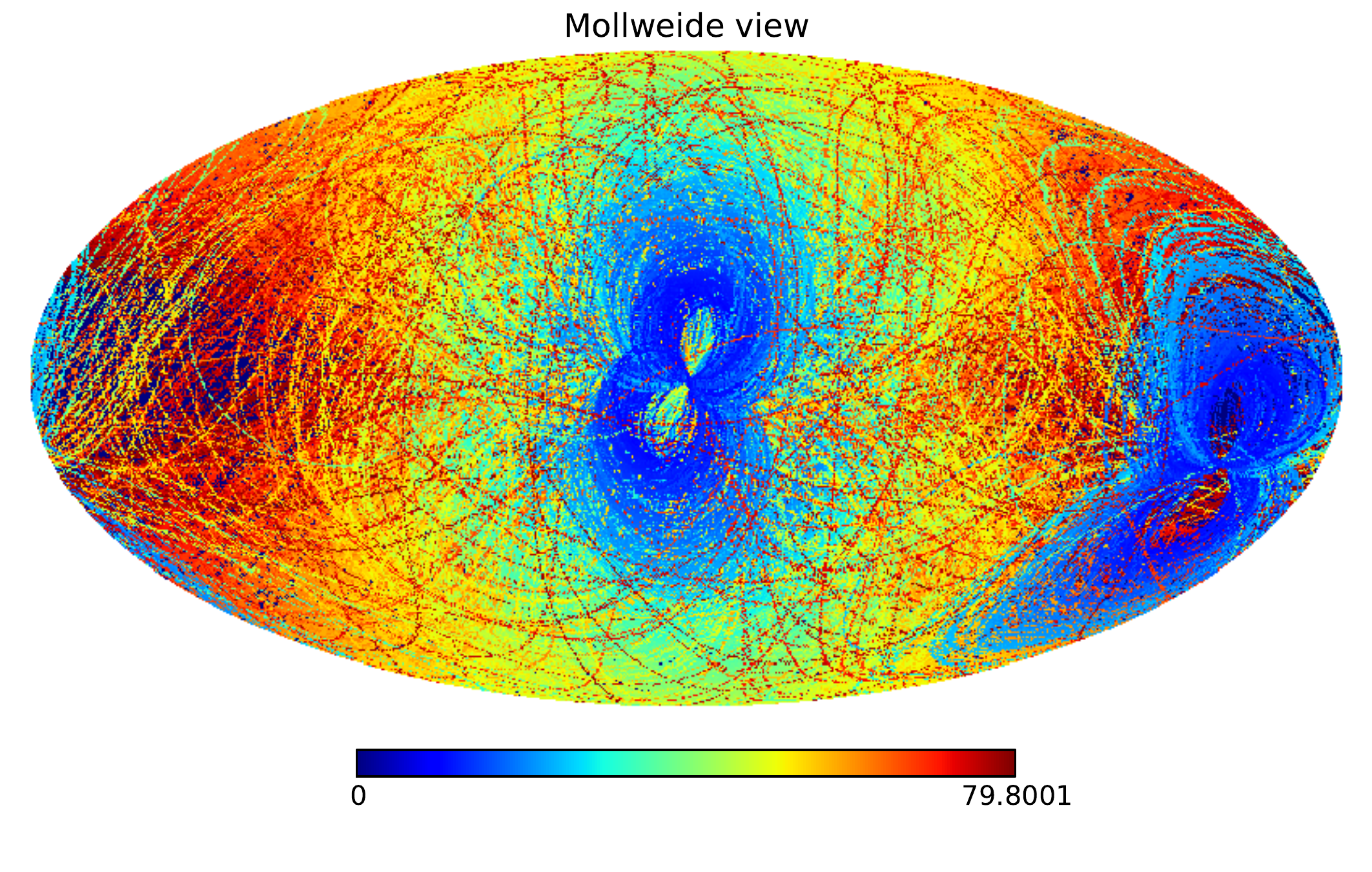}\\
  c)\includegraphics[width=0.45\linewidth]{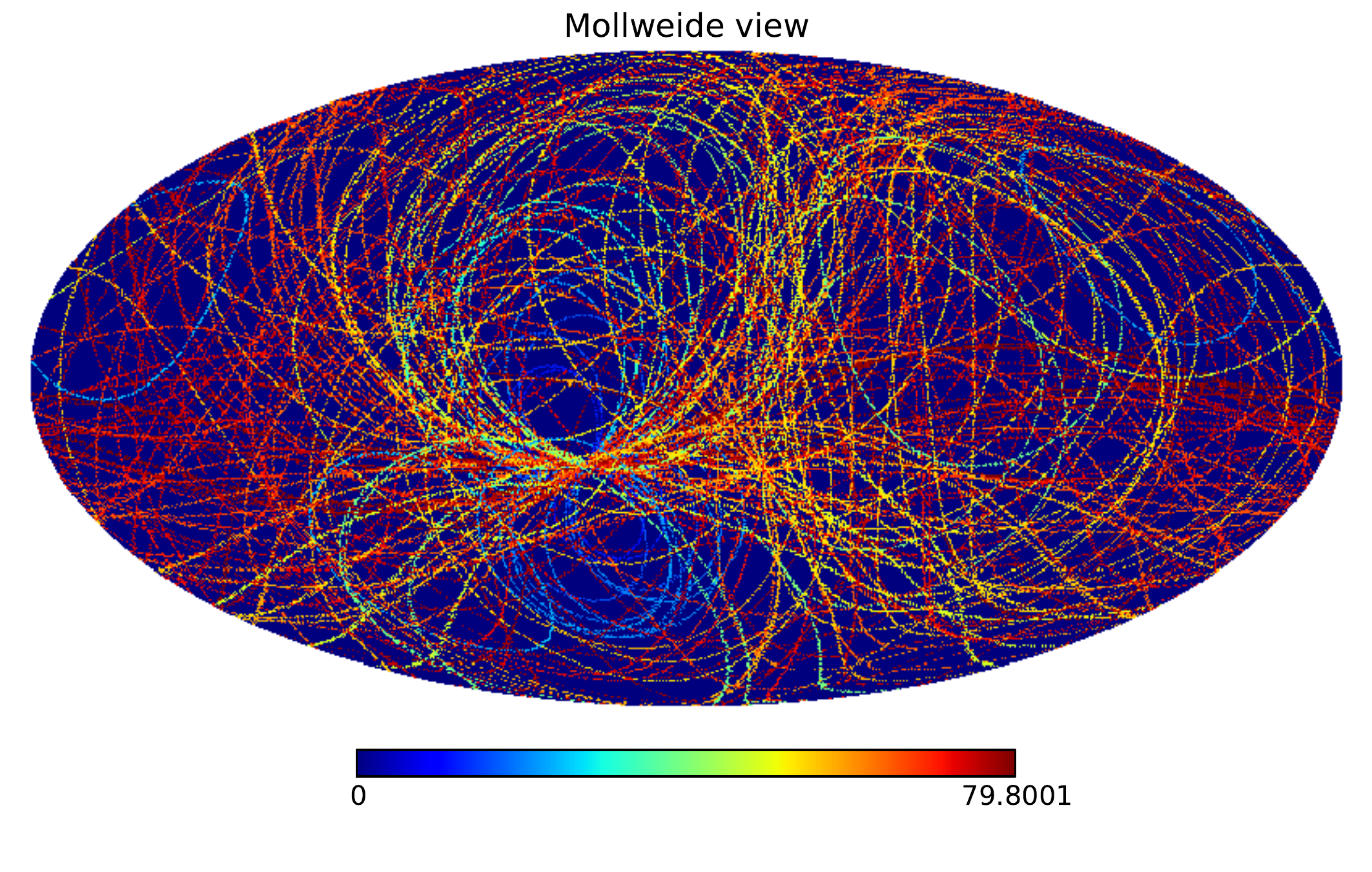}
  d)\includegraphics[width=0.45\linewidth]{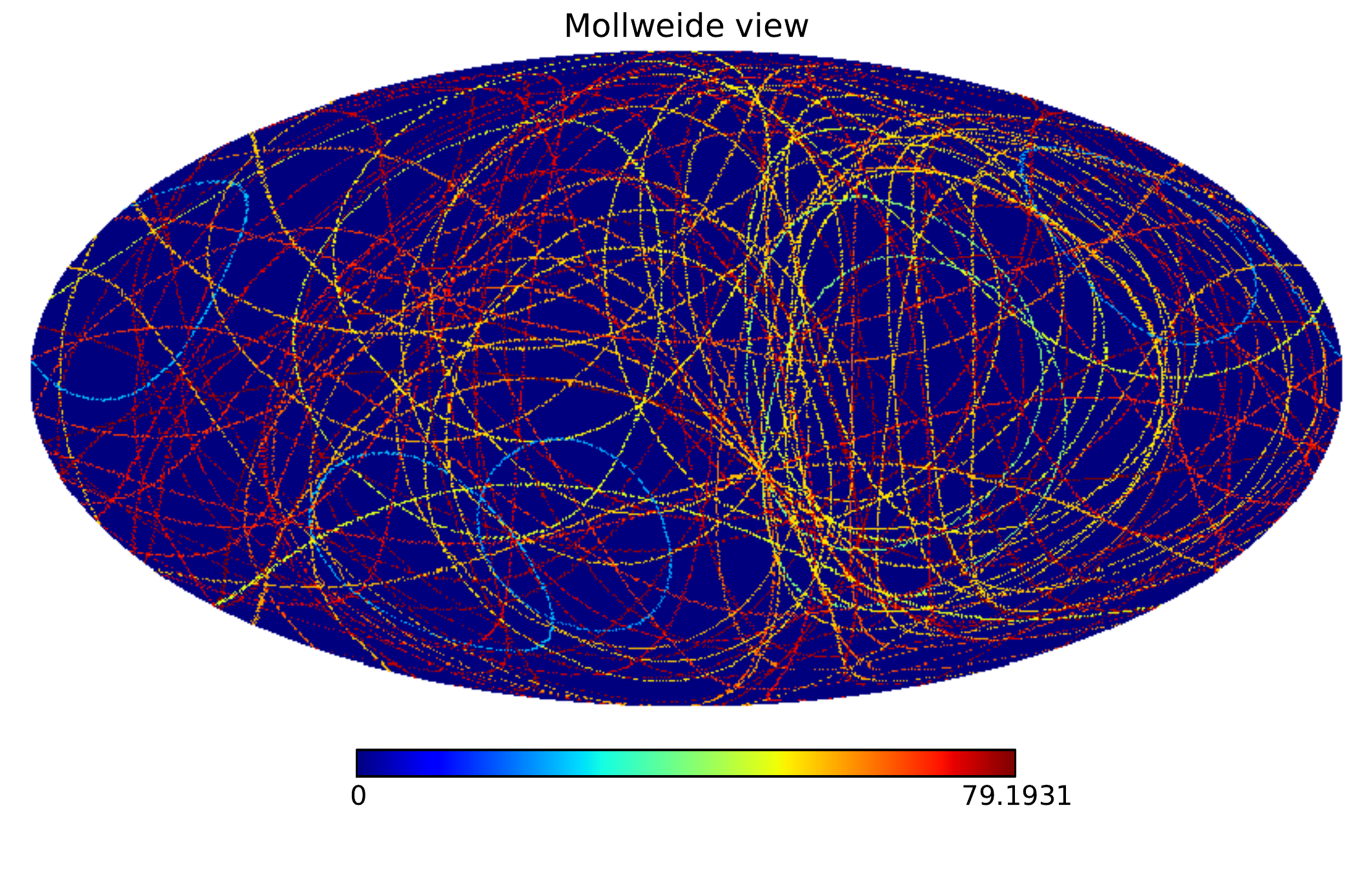}
  \caption{For the search for non-orientable manifolds: Location of
    the circles pairs with maximal statistics $S$ that lies above the
    $3\sigma$ CL, colored by opening angle $\alpha$. Note that the
    circles that touch either the galactic anti-center or
    $l=109.44^\circ$, $b=27.8^\circ$ do not hint at a non-trivial
    topology. a) The highest signal comes from circle pairs osculating
    at the galactic anti-center and at $l=109.44^\circ$,
    $b=27.8^\circ$. b) same as a), but rotated by $180^\circ$ such
    that the galactic anti-center is in the middle of the plot. c)
    Removing circle pairs that touch the galactic anti-center, the
    highest signal comes from pairs that touch $l=109.44^\circ$,
    $b=27.8^\circ$. d) Removing circles that touch either the galactic
    anti-center or $l=109.44^\circ$, $b=27.8^\circ$, no special
    position on the sky is apparent.}
  \label{fig:w_non_orientable_skymap}
\end{figure*}
The source of these excursions are again circles which touch either of
two distinct spots: the galactic anti-center and $l=109.44^\circ$,
$b=27.8^\circ$ (see
Figure~\ref{fig:w_non_orientable_skymap}a). Disregarding all circles
that come within $2.5^\circ$ of the galactic anti-center, we still
find many excursions above the $3\sigma$ CL (see
Figure~\ref{fig:w_non_orientable_skymap}b).  Finally, also removing
circles that come within $2.5^\circ$ of the position $l=109.44^\circ$,
$b=27.8^\circ$, there is no structure left (see
Figures~\ref{fig:w_non_orientable_skymap}c and
\ref{fig:w_non_orientable_exclude_galactic_anticenter2_5_exclude_special_position2_5b}).
\begin{figure*}
  \includegraphics[width=\figsize\linewidth]{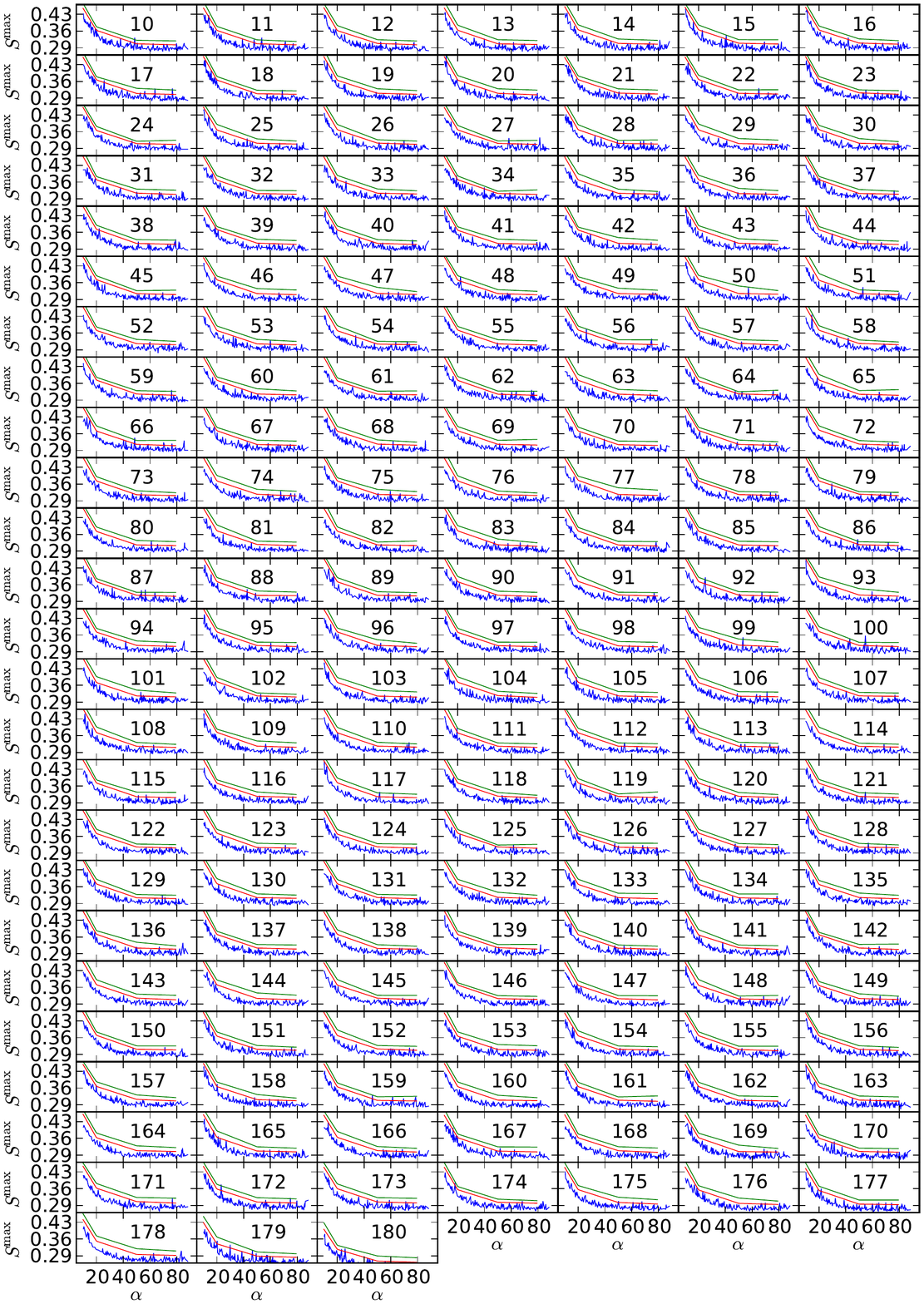}
  \caption{Searching for non-orientable topologies, final result: The
    statistics $S$ (solid blue line) as defined in
    Equation~\ref{eq:def:smax} as a function of opening angle $\alpha$
    and separation angle $\theta$ (indicated by the numbers in the
    center of each plot). The red/ green line correspond to the
    $2\sigma/ 3\sigma$ CL. We disregard circles that touch either the
    galactic anti-center or $l=109.44^\circ$, $b=27.8^\circ$. No signs
    of a non-trivial non-orientable topology are found.}
  \label{fig:w_non_orientable_exclude_galactic_anticenter2_5_exclude_special_position2_5b}
\end{figure*}

Figure~\ref{fig:w_non_orientable_exclude_galactic_anticenter2_5_exclude_special_position2_5b}
presents the final, negative result of the search for a non-trivial,
non-orientable topology of the universe. The remaining excursions
above $2\sigma$ and $3\sigma$ are well explained by random statistical
fluctuations as outlined in the previous subsection.  This leads us to
conclude that the data is consistent at 99.7\% CL with the null
hypothesis of no non-trivial, non-orientable topology.

\section{Conclusions and Outlook}

We employed the circles-in-the-sky statistics first devised in
\cite{Cornish:2003db}, looking for pairs of matching circles of
opening angles $10^\circ<\alpha<90^\circ$ and separation angles
$11^\circ\le\theta\le180^\circ$. We positioned the circle centers on a
grid with $N_\text{side}=128$, but computed the statistics on the full
$N_\text{side}=512$ CMB map.

While the WMAP 7 year data brought quite some improvements in the
noise of the $\Smax$ statistics (c.f. Figure 2 in
\cite{Cornish:2003db}), we find no hints of a non-trivial topology of
the universe (see
Figures~\ref{fig:w_orientable_exclude_galactic_anticenter2_5_exclude_special_position2_5}
and
\ref{fig:w_non_orientable_exclude_galactic_anticenter2_5_exclude_special_position2_5b}).
The new search covered a much wider range of possible topologies, and
by extending the search to circles with opening angles as small as
$10^\circ$, we have extended the previous bound on the size of the
Universe to $98.5\%$ of the diameter of the last scattering surface,
or approximately $26$Gpc.

There are systematic effects coming from both members of a circle pair
touching either the galactic anti-center or the position
$l=109.44^\circ$, $b=27.8^\circ$ (see
Figures~\ref{fig:w_orientable_skymap} and
\ref{fig:w_non_orientable_skymap}). As these positions appear both
when looking for orientable and non-orientable manifolds, they cannot
be of topological origin, but point towards a contamination of the map
at these positions.  The galactic anti-center region contains
significant amounts of galactic emission.  While the ILC maps used in
this analysis attempt to remove most of this emission, the correlated
residuals are a likely source of contamination in the circle searches.

We are looking forward to the data release of the Planck mission,
which will offer an exciting new, sharper view of the surface of last
scattering, allowing for a better search of signs of non-trivial
topology by removing noise particularly at smaller separation angles
$\alpha<30^\circ$. Further advances in computing power will enable a
search on a full $N_\text{side}\ge 512$ grid of circle positions.

\section*{Acknowledgements} 
It is a pleasure to thank Jeff Weeks for interesting discussions. Some
of the results in this paper have been derived using the
HEALPix\cite{Gorski:2004by} package. This work made use of the High
Performance Computing Resource in the Core Facility for Advanced
Research Computing at Case Western Reserve University. We acknowledge
the use of the Legacy Archive for Microwave Background Data Analysis
(LAMBDA). This work was supported in part by an allocation of
computing time from the Ohio Supercomputer Center. Support for LAMBDA
is provided by the NASA Office of Space Science. GDS and PMV were
supported by a grant from the US Department of Energy to the particle
astrophysics theory group at CWRU.  PMV was also supported by the
office of the Dean of the College of Arts and Sciences, CWRU, and the
Impuls und Vernetzungsfond of the Helmholtz Association of German
Research Centers under grant HZ-NG-603, and German Science Foundation
(DFG) within the Collaborative Research Center 676 “Particles, Strings
and the Early Universe”.

\bibliographystyle{unsrt}
\bibliography{circles}
\end{document}